\documentclass[aps,prb,twocolumn,showpacs]{revtex4}
\usepackage{graphicx}

\begin{document}

\title{Possible superconducting fluctuation and pseudogap state above $T_c$ in CsFe$_2$As$_2$}

\author{Huan Yang$^{\dag}$, Jie Xing, Zengyi Du, Xiong Yang, Hai Lin, Delong Fang, Xiyu Zhu, and Hai-Hu Wen$^{*}$}

\affiliation{National Laboratory of Solid State Microstructures, Center for Superconducting Physics and Materials, School of Physics, Collaborative Innovation Center for Advanced Microstructures, Nanjing University, Nanjing 210093, China}

\begin{abstract}
 Resistive, magnetization, torque, specific heat and scanning tunneling microscopy measurements are carried out on the hole heavily doped CsFe$_2$As$_2$ single crystals. A characteristic temperature $T^*\sim13$ K, which is several times higher than the superconducting transition temperature $T_c=2.15$ K, is observed and possibly related to the superconducting fluctuation or the pseudogap state. A diamagnetic signal detected by torque measurements starts from the superconducting state, keeps finite and vanishes gradually until a temperature near $T^*$. Temperature dependent resistivity and specific heat also show kinks near $T^*$. An asymmetric gap-like feature with the energy of 8.4 meV and a symmetric superconducting related gap of 2.2 meV on the scanning tunneling spectra are detected, and these pseudogap-related features disappear at temperatures up to at least 9 K. These observations by different experimental tools suggest the possible existence of superconducting fluctuation or pseudogap state in the temperature range up to 4 - 6 times of $T_c$ in CsFe$_2$As$_2$.
\end{abstract}

\maketitle

\section{Introduction}

The iron-based superconductors and cuprates are the only two unconventional systems with high temperature superconductivity. Superconductivity (SC) can be achieved by doping electrons or holes to the parent compounds in these two systems when the antiferromagnetic order is suppressed, so these two systems have rather similar doping phase diagram\cite{PhaseDiagram,WenReview}. The pseudogap (PG) is an important feature on the phase diagram of cuprates, and has attracted many experimental and theoretical attentions\cite{PGReview}. The relationship between PG and SC in cuprates is still under debate\cite{PGReview,PGReview2} which challenges the basic description of the Landau Fermi liquid. A recent scanning tunneling microscopy (STM) study suggests that PG in Bi$_2$Sr$_2$CaCu$_2$O$_{8-\delta}$ is the intrinsic property of charge reservoir BiO layers, and has very little relation to SC\cite{PG2212}. PG is not quite common, but also observed in several families of iron-based superconductors from different measurements\cite{PGFeLaOF,PGFeSmOF,PGBaKFeAs,PGBaKFeAs2,PGBaFeCoAs,PGNaFeCoAs,PGBaFeAsP}, and even has the similar doping dependent behavior in Co doped iron pnictides as that in cuprates\cite{PGBaFeCoAs,PGBaFeCoAs2,PGBaFeCoAs3}. Superconducting fluctuation (SCF) is another important property in cuprates, which was proved by the Nernst signal and diamagnetic magnetization far above $T_c$ \cite{CuNernstReview}. In contrast, the temperature range of SCF in iron pnictides seems not so wide\cite{SCFLaOFeAs,SCFBaFeCoAs,SCFFeSeTe,SCFBaFeAsP}.

Although there is very rich physics in the two high-$T_c$ systems as mentioned above, the superconducting mechanism has not been settled yet. For the gap symmetry, cuprates have the well-known and dominant $d$-wave gap, while the situation in iron-based superconductors is very complex because Fermi-surface topologies possess tremendous difference in different materials\cite{FSTopo}. A widely accepted pairing symmetry for some iron pnictides is the $s_{\pm}$ pairing manner which needs the nesting condition between the hole and electron pockets with almost equal sizes\cite{s+-}, and this picture gets some support from the STM\cite{HanaguriSTM,CuImpurity} and other measurements. AFe$_2$As$_2$ (A = K, Cs, Rb) is in the extremely hole-doping level of the 122 family in iron pnictides. It is assumed that the correlation effect is getting more and more strong starting from the parent phase BaFe$_2$As$_2$ to the heavily hole doped case in AFe$_2$As$_2$ (A = K, Cs, Rb). The pairing symmetry of these materials may be different from the early proposed $s_{\pm}$ since they have have only the hole pockets. A nodal superconducting gap was suggested in AFe$_2$As$_2$ by different methods\cite{PenDepthKFeAs,ShinKFeAs,TailleferKFeAs,LiSYCsFeAs,LiSYRbFeAs}. Our recent work on KFe$_2$As$_2$ reports a Van Hove singularity just several meV below the Fermi level ($E_F$), and it have essential influence on the SC in the material\cite{VanHove}. $T_c$ of AFe$_2$As$_2$ family under a pressure less than 3.3 GPa shows a universal V-shaped phase diagram\cite{KFeAsHP,AFeAsHP}. Another report shows two separate superconducting region under high pressure up to 33 GPa in KFe$_2$As$_2$, and $T_c$ suddenly jumps to about 11 K at 14.4 GPa in the second superconducting phase followed by a sign change of the Hall coefficient\cite{KFeAsHP2}. All these indicate that the physics in the heavily hole doped systems AFe$_2$As$_2$ (A = K, Cs, Rb) may be  more complex than the light doped systems. This inspires us to investigate the rich physics of these systems with different experimental tools. In this paper, we report experimental results on CsFe$_2$As$_2$ single crystals from multiple measurements. We find a clear characteristic temperature $T^*\sim 13$ K which may correspond to the superconducting fluctuation or the pseudogap state temperature and is much higher than $T_c$. This phenomenon adds very interesting new ingredients in understanding the correlation effect and superconductivity in these complex systems.

\section{Experiments}

The CsFe$_2$As$_2$ single crystals were synthesized by using the self-flux method\cite{selfflux}. The Cs chunks, Fe and As powders were weighted by the ratio of Cs:Fe:As=6:1:6. The mixture was put into an alumina crucible which was sealed in a tantalum crucible in high-purity argon atmosphere. The crucible sealed in a evacuated quartz tube was slowly heated up to 200$^\circ$C and held for 6 h afterwards. It was then heated to 950$^\circ$C and held for 10 h, cooled down to 550$^\circ$C at the rate of 3$^\circ$C/h to grow single crystals. The sample was cooled down to room temperature by shutting off the power of the furnace finally. Shiny plate-like single crystals can be obtained from black CsAs flux. The extra flux on the surface of the crystal can be effectively washed out by deion water or alcohol. The magnetization measurements were carried out by a superconducting quantum interference device system SQUID-VSM (Quantum Design), while the resistance, specific heat, and torque data were measured by a physical property measurement system PPMS-16 (Quantum Design). The scanning tunneling microscopy/spectroscopy (STM/STS) measurements were done in a ultrahigh vacuum, low temperature, and high magnetic-field scanning probe microscope USM-1300 (Unisoku Co., Ltd.).

\section{Results}

\begin{figure}
\includegraphics[width=8cm]{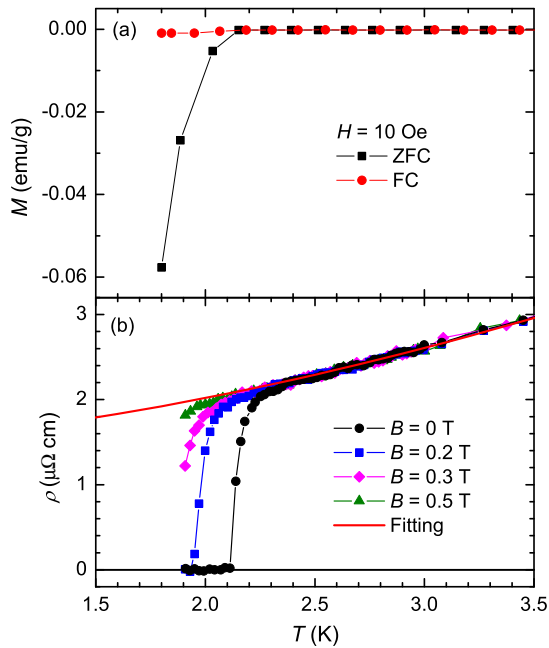}
\caption{(Color online) (a) Temperature dependence of magnetization in CsFe$_2$As$_2$ after ZFC and FC processes. (b) Temperature dependent resistivity at different magnetic fields. The red line is a power-function fitting result to the experimental resistive curve measured at zero-field from 2.4 K to 10 K.} \label{figTc}
\end{figure}

Figure~\ref{figTc}(a) shows the temperature dependence of the mass magnetization (M) after zero-field-cooling (ZFC) and field cooling (FC) processes at 10 Oe. Figure~\ref{figTc}(b) shows temperature dependence of resistivity at different fields with the lowest temperature to 1.9 K, by which one can determine the zero-resistance transition temperature $T_{c0}=2.11$ K at 0 T. Then we use a power function $\rho(T)=\rho_0+AT^n$ to fit the experimental resistive curve measured at 0 T from 2.4 K to 10 K, the fitting result with $\rho_0=1.43\ \mu\Omega\cdot$cm and $n=1.7$ is shown as a solid red line in Fig.~\ref{figTc}(b). Compared with the resistivity value 590 $\mu\Omega\cdot$cm of this sample measured at 300 K (data not shown here), we can obtain the residual resistance ratio $RRR\equiv \rho\left(T=300\ \mathrm{K}\right)/\rho_0=413$ which is much smaller than that reported in KFe$_2$As$_2$ \cite{KFe2As2} and larger than other ratios from previous reports\cite{LiSYCsFeAs,ChenXHCsFeAs}. The zero-field resistive curve shows a very narrow superconducting fluctuating region with the width of about 0.1 K near the onset transition temperature if we use the red fitting curve as the normal state resistivity. The sample is still in the superconducting state at 1.9 K and 0.2 T from the resistive curve, while the superconductivity has already become very weak when the field increases to 0.5 T at the same temperature.

\begin{figure}
\includegraphics[width=8cm]{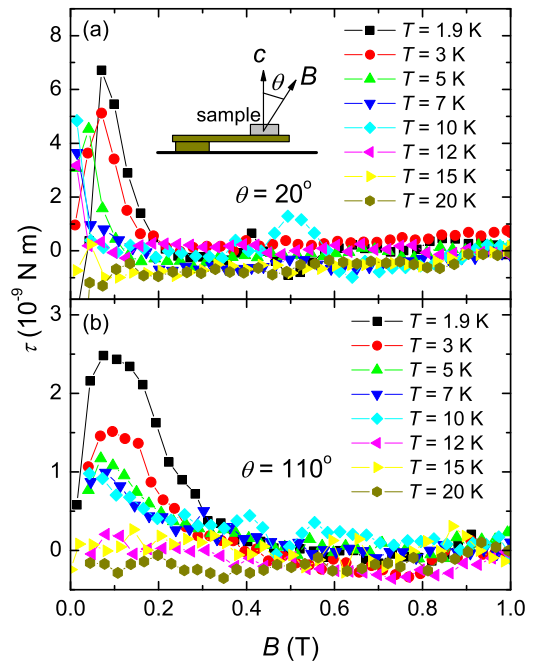}
\caption{(Color online) Field dependent of torque signal measured at different temperatures with the angle between the magnetic field and $c$-axis of the sample $\theta=20^\circ$ for (a) and $\theta=110^\circ$ for (b). The inset in (a) shows the measurement configuration and the definition of $\theta$.}
\label{figTorque}
\end{figure}

A diamagnetic magnetization signal above $T_c$ measured in cuprates by torque magnetometry is shown to be an efficacious proof of the superconducting fluctuation beside the Nernst signal\cite{TorqueBi2212}. The data of magnetic field $B$ dependent torque signal measured at different temperatures are shown in Fig.~\ref{figTorque}. The angle $\theta$ between the magnetic field and $c$-axis of the sample [shown in the inset of Fig.~\ref{figTorque}(a)] are $20^\circ$ and $110^\circ$ for Fig.~\ref{figTorque}(a) and Fig.~\ref{figTorque}(b), respectively. The two selected angles correspond to the positions for maximum and minimum values in angular dependent torque magnitude measured at 1.9 K and 0.2 T. The measured torque can be expressed as $\vec{\tau}= \vec{m}\times \vec{B}$, where $\vec{m}$ is the magnetic moment of the sample. In the superconducting sates, the diamagnetic magnetization should be a function of $B$ as performing in the magnetization curve with increasing of $B$ from zero-field. As the product of the magnetization and the magnetic field, the torque signal should begin from zero at zero magnetic field and then reach the maximum following by a decrease to zero when the magnetic field destroys the superconductivity. The existence of the extremum values in angular dependent torque suggests the anisotropic magnetization of the sample in the superconducting state. The sample is obviously in its superconducting state when $B<0.5$ T at 1.9 K from Fig.~\ref{figTc}(b), so the torque signal measured in the same condition is from the superconducting diamagnetic effect. The torque signal peak always exists and doesn't change its sign when the temperature is increased crossing the critical temperature of 2.15 K, but the magnitude of the the signal will reduce and finally disappear to the background above 12 K for both two configurations. The sample should be very tiny for the torque measurement, and the magnetization is also weakened by increasing of magnetic field. So it is difficult to judge the exact magnetic field at which the diamagnetic signal disappears. The field $B_c^\tau$ at which the torque signal merges to the background in Fig.~\ref{figTorque}(a) and (b) at 1.9 K is about 0.2 T ($\theta=20^\circ$) or 0.4 T ($\theta=110^\circ$) in our resolution. Here from Fig.~\ref{figTorque}, we can find $B_c^\tau$ has a temperature dependent behavior and anisotropy for the two angles, which is another proof of the signal from the superconducting diamagnetic effect.

\begin{figure}
\includegraphics[width=8cm]{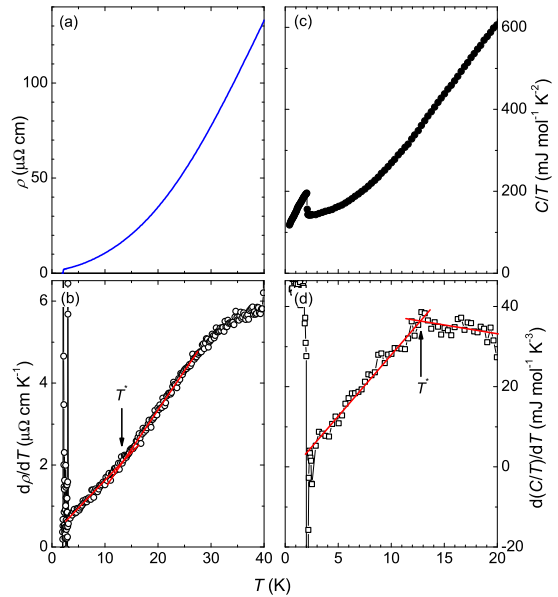}
\caption{(Color online) Temperature dependence of resistivity (a) and specific heat (c) and the corresponding differential curves (b) and (d) in a wide temperature range. One can find the kink behaviors in both differential curves at temperature about $T^*=13$ K denoted by the arrows. The red solid lines are guides for eyes.} \label{figRCap}
\end{figure}

We have observed that a diamagnetic signal disappeared above 12 K in torque measurement, however the fluctuation seems not so strong from the resistive measurement near the transition as shown in Fig.~\ref{figTc}(b). Then we analyze the resistivity data in a wider temperature scale in Fig.~\ref{figRCap}(a), and calculate the temperature derivative to the resistive curve as shown in Fig.~\ref{figRCap}(b). One can find a kink at $T^*=13$ K and the slop changes a bit on its both sides. Heat capacity is a very sensitive method to detect bulk properties of phase transitions. The superconducting transition can be observed very well in Fig.~\ref{figRCap}(c), which is similar to the previous report\cite{ChenXHCsFeAs}. One can find a obvious turning point near $T^*=13$ K from the temperature dependent differential heat capacity in Fig.~\ref{figRCap}(d). Therefore one can draw a self-conclusion that there is a characteristic temperature at which both the resistivity and the heat capacity shows a kink in their temperature derivative curves. This temperature is also near the one that torque signal disappears. We argue that this may correspond to the transition for the possible superconducting fluctuation, and the transition temperature is about 6 times of $T_c$.

\begin{figure}
\includegraphics[width=8cm]{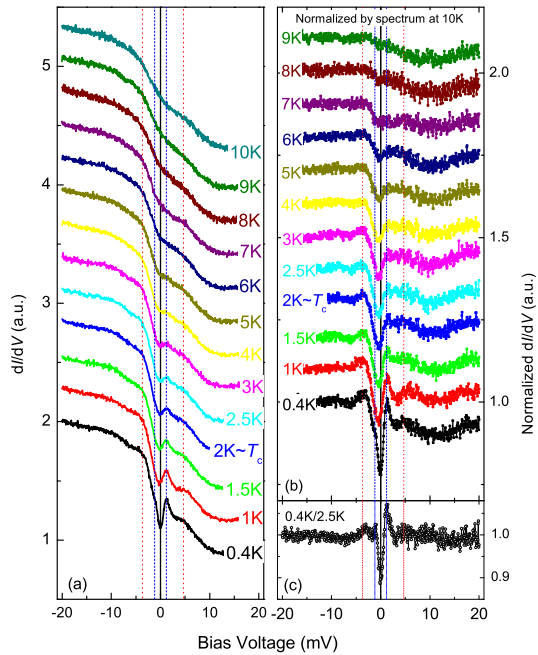}
\caption{(Color online) (a) The evolution of the STS spectra measured at temperatures from 0.4 K to 10 K. (b) STS spectra normalized by the one measured at 10 K. (c) STS spectrum measured at 0.4 K normalized by the one measured at 2.5 K just above the bulk $T_c$. Besides the superconducting gap peaks near $\pm1.2$ mV, there is another pair of asymmetric gap features locating at $-3.7$ and 4.7 mV on the spectra. }
\label{figSTS}
\end{figure}

We also carried out the STS measurements on the CsFe$_2$As$_2$ single crystals to detect the local electronic properties. A typical set of STS spectra measured at different temperatures is shown in Fig.~\ref{figSTS}. In KFe$_2$As$_2$, a Van Hove singularity was detected locating just a few meV below $E_F$, and it generates a strong peak of DOS on the STS spectra in a wide energy range of dozens of meV\cite{VanHove}. However, there is no obvious peak near $E_F$ on the spectra in CsFe$_2$As$_2$, which suggests that the possible Van Hove singularity is far away from the Fermi energy due to the subtle change of the band structures between KFe$_2$As$_2$ and CsFe$_2$As$_2$. However, the spectrum in CsFe$_2$As$_2$ has an asymmetric background for the two sides above and below the Fermi level, which is similar to that in KFe$_2$As$_2$ and consistent with energy dependent density of states (DOS) of AFe$_2$As$_2$ from theoretical calculations\cite{AFeAsTheo}. From the spectrum measured at 0.4 K normalized by the one measured at 2.5 K as shown in Fig.~\ref{figSTS}(c), we can find a superconducting gap with the peak-peak distance of $2\times1.2$ meV which is comparable with that in KFe$_2$As$_2$. However the real superconducting gap value depends on the detailed fitting with gap symmetry of the sample. There is also a quite high ungapped DOS on the Fermi level, which is similar to the situation in KFe$_2$As$_2$. Besides the symmetric coherence peaks of the sample, there are also two asymmetric gap features with the peak position near $-3.7$ and 4.7 mV as shown by the dotted lines in Fig.~\ref{figSTS}(b,c). This asymmetric feature may be induced by the asymmetric background of the DOS on energy. It should be noted that the coherence-peak like feature almost disappears at 2.5 K above the bulk $T_c$, but the asymmetric large-gap feature near $E_F$ extends to at least 9 K. The gap feature is not observable above 10 K partially because of the thermal broadening effect. It is difficult to observe any other gap features on KFe$_2$As$_2$ because of the strong peak near $E_F$ from Van Hove singularity, but we can also find some trace of the dip feature at $E_F$ on the spectra measured just above $T_c$\cite{VanHove}. Since the bulk onset superconducting temperature is 2.4 K in CsFe$_2$As$_2$ as shown in Fig.~\ref{figTc}(b), the two-gap feature above $T_c$ may be related to the pseudogap or the superconducting fluctuation mentioned above.

\section{Discussion}

In the cuprates, the pseudogap is regarded as a symmetric gap directly developed from the superconducting gap above $T_c$, and getting support from the bulk measurements, e.g., resistivity, nuclear magnetic resonance, specific heat, Nernst effect, etc. \cite{PGReview,PGReview2}. From our STS data, we find two energy gaps, i.e., one smaller gap $2\Delta_1=2.2$ meV closely related to the SC with symmetric peak energies to $E_F$ and the other $2\Delta_2=8.4$ meV with asymmetric peak energies with respect to $E_F$. In the iron pnictides, the asymmetric pseudogap-like feature is observed above $T_c$ on Na(Fe$_{1-x}$Co$_x$)As \cite{PGNaFeCoAs}, so we cannot exclude the possibility of the asymmetric gap feature as the pseudogap. Since there is 80\% - 90\% ungapped DOS near $E_F$ and asymmetric background of DOS in the normal state, it is very difficult to recognize which gap related to the possible pseudogap in the material. Both gaps in addition with the dip near $E_F$ disappear above 9 K, and it is also difficult to determine the ending temperature because of the thermal broadening effect. We have found self-consistent support from the bulk resistance and specific heat measurement from which a characteristic temperature $T^*\sim13$ K is observed. Here with the two possible pseudogap values, one can obtain the reduced gap ratios $2\Delta_1/k_BT_c=11.6$ and $2\Delta_2/k_BT_c=45$ which are both in the range of 9 - 46 summarized in the ``1111''-family iron-based superconductors\cite{PGrange}.

One of the early understandings for the pseudogap in cuprates is its close relationship to Cooper pairing above $T_c$, however critical temperatures for superconducting fluctuation and pseudogap state are divorced from each other by a careful STS measurement\cite{PGSCF}. The pseudogap in normal state is also regarded as a consequence of phase-incoherent Cooper pairs in iron-based superconductors\cite{SCFBaFeCoAs}. Recently, it was concluded that superconducting fluctuation may be very strong in the FeSe single crystals because of the vicinity to the BEC-BCS crossover\cite{MatsudaPNAS}. This recommends a picture of preformed Cooper pairs with phase incoherence. In our experiments, the diamagnetic signal was detected by torque measurements up to more than 12 K, and the critical field $H_c^\tau$ above $T_c$ has a temperature dependence as well as an anisotropy to the angle between the magnetic field and $c$-axis of the crystal. The characteristic temperature derived from different experimental tools (beside that from resistivity) is also near $T^*\sim6 T_c$. Since the excess conductivity due to the possible superconducting fluctuation is not observed, we thus can not conclude that the pseudogap effect is due to the superconducting fluctuation. As addressed in early transport measurements by resistivity and Nernst in cuprates, the resistivity is less pronounced to illustrate the superconducting fluctuation compared with the Nernst signal. Therefore a Nernst measurement is highly desired although the temperature is very low and the window is quite narrow. However either pseudogap feature or superconducting fluctuation in present system should be closely related to the superconductivity mechanism, which may suggest a strong pairing gap, but weak phase coherence in CsFe$_2$As$_2$.

\section{Concluding Remarks}

By using the torque measurement, we have observed the diamagnetic signal at temperatures of 6 times of $T_c$ in CsFe$_2$As$_2$ single crystals. This may be explained by the superconducting fluctuation effect. Supporting evidence is obtained through analyzing the derivative of resistivity and specific heat, both illustrate a characteristic temperature $T^*\sim13$ K. In addition, the pseudogap feature is detected as the appearance of a second set of larger gaps existing until 9 K on the tunneling spectra. All these facts point to a convergent picture that there is a pseudogap effect in CsFe$_2$As$_2$. Our results present a picture of strong pairing with weak phase coherence in the extremely hole doped system, which will provide extra information to the superconducting mechanism in iron based superconductors.

\section*{Acknowledgments}
This work is supported by the NSF of China, the Ministry of Science and Technology of China (973 Projects: No. 2011CBA00102, 2012CB821403).

$^{\dag}$ huanyang@nju.edu.cn
$^*$ hhwen@nju.edu.cn

\end{document}